\documentstyle[psfig]{mn}

\def\ga{\mathrel{\hbox{\rlap{\hbox{\lower4pt\hbox{$\sim$}}}\hbox{$>$}}}}
\def\fd{\hbox{$.\!\!^{\rm d}$}}

\title[High-speed photometry of faint Cataclysmic Variables: II.]
{High speed photometry of faint Cataclysmic Variables: II. RS Car, V365 Car, V436 Car, AP Cru,
RR Cha, BI Ori, CM Phe and V522 Sgr}
\author[Patrick A. Woudt and Brian Warner]
       {Patrick A. Woudt\thanks{E-mail: pwoudt@circinus.ast.uct.ac.za} 
        and Brian Warner\thanks{E-mail: warner@physci.uct.ac.za}\\
        Department of Astronomy, University of Cape Town, Private Bag,
        Rondebosch 7700, South Africa}
\date{}

\begin{document}

\maketitle

\begin{abstract}
Short time scale photometric properties of eight faint Cataclysmic Variable (CV) stars are presented.
Nova Carinae 1895 (RS Car) has a photometric modulation at 1.977 h which could be either an orbital
or a superhump period. Nova Carinae 1948 (V365 Car) shows flickering, but any orbital modulation has a
period in excess of 6 h. The nova-like variable and X-ray source V436 Car has an orbital modulation at 
$P_{orb}$ = 4.207 h, no detectable period near 2.67 h (which had previously given it a possible
intermediate polar classification), and Dwarf Nova Oscillations (DNOs) at $\sim$40 s. Nova Crucis 1936
(AP Cru) has a double humped ellipsoidal modulation at $P_{orb}$ = 5.12 h and a stable modulation
at 1837 s characteristic of an intermediate polar. Nova Chamaeleontis 1953 (RR Cha) is an ecliping
system with $P_{orb}$ = 3.362 h, but at times shows negative superhumps at 3.271 h and positive
superhumps at 3.466 h. In addition it has a stable period at 1950 s, characteristic of an intermediate polar.
BI Ori is a dwarf nova which we observed at quiescence and outburst without detecting any orbital modulation. 
CM Phe is a nova-like variable for which we confirm Hoard, Wachter \& Kim-Quijano's (2001) value of $P_{orb}$ = 6.454 h.
We have identified the remnant of Nova Saggitarii 1931 (V522 Sgr) with a flickering source $\sim$2.2 mag
fainter than the previously proposed candidate (which we find to be non-variable).
\end{abstract}


\section{Introduction}

The first paper in this series (Woudt \& Warner 2001, Paper I) explained the philosophy
driving our survey of faint Cataclysmic Variable stars (CVs). In essence, the combinations of basic
parameters (masses, mass transfer rates, periods, magnetic fields) and viewing angles (orbital inclination) result
in no two CVs appearing alike. Furthermore, there has been a steady increase in revealed new phenomena as
larger numbers of CVs have been studied in detail. Also, the advent of 8-m class telescopes opens opportunities for
high resolution studies of much fainter CVs than hitherto, so a preliminary examination of such stars down to
magnitudes 20 or 21 is now justified in order to discover the more interesting specimens.

The observations reported here were in almost all cases made with the University of Cape Town (UCT) CCD 
photometer (O'Donoghue 1995) attached to the 1.0-m (40-in) or 1.9-m (74-in) reflectors at the Sutherland site 
of the South African Astronomical Observatory. As in Paper I, the full wavelength response of the CCD was used, i.e.
we observed in white light. Our photometry was calibrated by observing hot white dwarf standards.
We have tended to concentrate on faint nova remnants in crowded fields, this having proved a rewarding area
for the discovery of objects of particular interest, and one where the UCT CCD photometer reveals its greatest
advantages.

The observations are given in Section 2 and some brief conclusions in Section 3. The list of our observations
is given in Table 1.

\begin{table*}
 \centering
  \caption{Observing log.}
  \begin{tabular}{@{}llrrrrrcc@{}}
 Object       & Type         & Run No.  & Date of obs.          & HJD of first obs. & Length    & $t_{in}$ & Tel. &  V \\
              &              &          & (start of night)      &  (+2451000.0)     & (h)       &     (s)   &      & (mag) \\[10pt]
{\bf RS Car}  & NR           & S6224    & 22 May 2001 & 1052.20050  &   6.47      &  10, 30   &  74-in & 18.5\\
              &              & S6226    & 23 May 2001 & 1053.19350  &   4.90      &      30   &  74-in & 18.4\\
              &              & S6235    & 28 May 2001 & 1058.21513  &   4.70      &      30   &  74-in & 18.5\\[5pt]
{\bf V365 Car}& NR           & S6068    &  8 Mar 2000 &  612.26342  &   3.01      &      10   &  74-in & 18.3\\
              &              & S6070    &  9 Mar 2000 &  613.25598  &   9.19      &      10   &  74-in & 18.3\\
              &              & S6073    & 11 Mar 2000 &  615.28591  &   8.60      &      20   &  74-in & 18.2\\[5pt]
{\bf V436 Car}& DN           & S6039    & 29 Dec 1999 &  542.41978  &   3.91      &   5, 10   &  40-in & 15.2\\     
              &              & S6040    & 30 Dec 1999 &  543.31581  &   6.46      &      10   &  40-in & 15.2\\     
              &              & S6041    & 31 Dec 1999 &  544.30565  &   6.39      &      10   &  40-in & 15.2\\     
              &              & S6044    &  3 Jan 2000 &  547.30356  &   6.69      &      10   &  40-in & 14.5\\[5pt]
{\bf RR Cha}  & NR           & S6195    & 24 Feb 2001 &  965.54565  &   1.97      &      60   &  40-in & 18.3\\
              &              & S6198    & 25 Feb 2001 &  966.45083  &   4.19      &      60   &  40-in & 18.2\\
              &              & S6200    & 26 Feb 2001 &  967.31425  &   7.35      &      60   &  40-in & 18.4\\
              &              & S6210    & 15 May 2001 & 1045.54546  &   3.23      &      60   &  40-in & 18.4\\
              &              & S6215    & 17 May 2001 & 1047.32779  &   8.53      &      60   &  40-in & 18.4\\[5pt]
{\bf AP Cru}  & NR           & S6069    &  8 Mar 2000 &  612.59236  &   1.21      &      10   &  74-in & 18.1\\
              &              & S6075    & 12 Mar 2000 &  616.33250  &   7.43      &      10   &  74-in & 18.1\\
              &              & S6077    & 13 Mar 2000 &  617.32908  &   6.09      &  10, 20   &  74-in & 18.0\\
              &              & S6094    &  1 Jun 2000 &  697.19578  &   2.87      &      10   &  74-in & 17.8\\[5pt]
{\bf BI Ori}  & DN           & S6253    & 24 Sep 2001 & 1177.52290  &   1.01      &       5   &  40-in & 14.9\\     
              &              & S6256    &  9 Oct 2001 & 1192.50259  &   0.54      &       6   &  40-in & 16.7\\     
              &              & S6261    & 10 Oct 2001 & 1193.49359  &   3.35      &       6   &  40-in & 16.4\\     
              &              & S6265    & 11 Oct 2001 & 1194.49802  &   0.73      &      30   &  40-in & 14.6\\     
              &              & S6275    & 15 Oct 2001 & 1198.49312  &   3.18      &       6   &  40-in & 15.6\\[5pt]
{\bf CM Phe}  & NL           & S6141    & 21 Dec 2000 &  900.28990  &   3.56      &      10   &  40-in & 15.1\\
              &              & S6145    & 22 Dec 2000 &  901.27896  &   2.63      &      10   &  40-in & 15.1\\
              &              & S6148    & 23 Dec 2000 &  902.27450  &   4.70      &      10   &  40-in & 15.2\\
              &              & S6151    & 24 Dec 2000 &  903.27397  &   3.31      &      10   &  40-in & 15.0\\
              &              & S6154    & 25 Dec 2000 &  904.27574  &   1.90      &      20   &  40-in & 15.2\\
              &              & S6177    & 29 Dec 2000 &  908.30416  &   2.71      &      20   &  40-in &  -- \\
              &              & S6180    & 30 Dec 2000 &  909.31926  &   2.52      &      20   &  40-in &  -- \\[5pt]
{\bf V522 Sgr}& NR           & S6101    &  5 Jun 2000 &  701.43325  &   1.66      &      15   &  74-in &19.7:\\[5pt]
\end{tabular}
{\footnotesize 
\newline 
Notes: NR = Nova Remnant, NL = Nova-like, DN = Dwarf Nova, $t_{in}$ is the integration time, `:' denotes an uncertain value.\hfill}
\label{tab1}
\end{table*}

\section{Observations}

\subsection{RS Carinae}

RS Car was Nova Carinae 1895, reached a photographic magnitude of at least $m_{pg} = 7.2$, and was probably 
a moderately fast nova. Its post-nova quiescent magnitude was thought to be $>$22 mag (Duerbeck 1987), but a recent
spectroscopic search has located it at $V \sim 18$, 7$''$ away from its previously estimated position (Bianchini et al.~2001).
The spectrum shows a blue continuum with a slope characteristic of an optically thick disc, on which are superimposed strong
HeII, NIII/CIII, and moderate Balmer emission lines. The H$\alpha$ equivalent width of 20 {\AA} measured by Bianchini
et al.~(2001), entered into the $W(H\alpha)$ -- inclination correlation diagram (Warner 1986), indicates an inclination
of $\sim$65$^{\circ}$.

\begin{figure*}
\centerline{\hbox{\psfig{figure=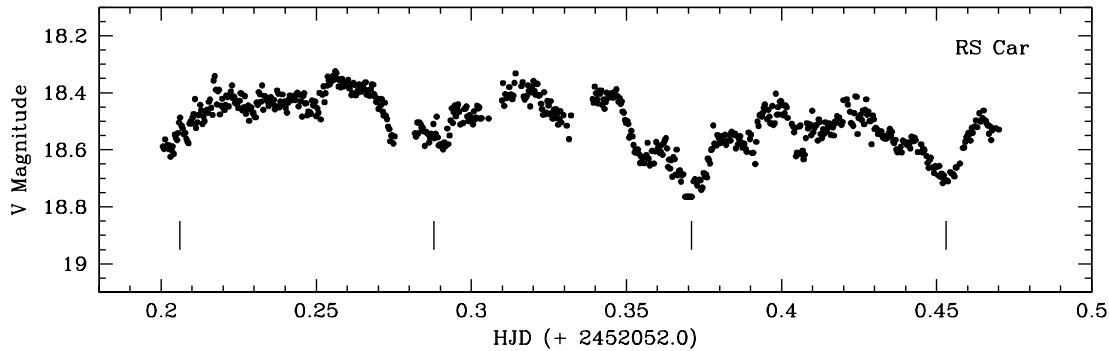,width=16cm}}}
  \caption{The light curve of RS Car on 2001 May 22.}
 \label{lcrscar}
\end{figure*}

\begin{figure}
\centerline{\hbox{\psfig{figure=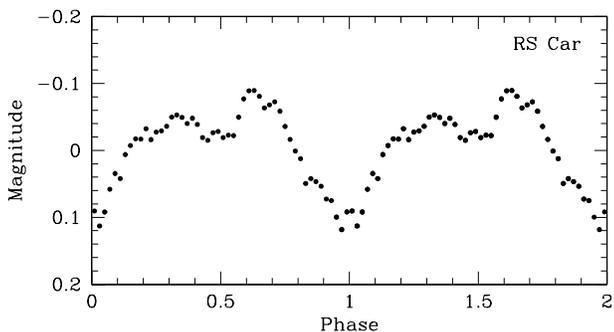,width=8.2cm}}}
  \caption{The average light curve of RS Car, folded on the 118.6 min period.}
 \label{rscar_av}
\end{figure}

The log of our photometric observations is included in Table 1. The longest of our light curves is displayed in Fig.~\ref{lcrscar}
and shows a clear modulation with a period near 2 h. From the Fourier transform of the entire data set we measure a period
of 1.977 h. The mean light curve at this period is shown in Fig.~\ref{rscar_av}. As expected from the profile, the Fourier
spectrum is rich in harmonics; the first, third and higher harmonics are quite strong but the second harmonic is very
weak. Prewhitening of the light curve at the fundamental and harmonics leaves no significant features.

We cannot tell from our short baseline whether the periodic brightness variation is an orbital or a superhump
modulation. From the gradient of the flux in the spectrum of RS Car, which resembles that of an optically thick 
accretion disc (Bianchini et al.~2001) and shows that the system is in the high $\dot{M}$ state characteristic of an
old nova, and the short orbital period (implying a low mass ratio), we expect the accretion disc to be 
permanently  perturbed into an eccentric shape. The period we have found is therefore probably a superhump period,
and the orbital period will be a few percent shorter. The profile of the light modulation is characteristic
of a superhump.

RS Car joins the small group of classical novae with orbital periods shorter than 2 h. The others are RW UMi ($P_{orb}$ 
= 1.419 h: Retter \& Lipkin 2001), GQ Mus ($P_{orb}$ = 1.425 h: Diaz et al.~1995), CP Pup ($P_{orb}$ = 1.474 h, $P_{sh}$ = 
1.500 h: Patterson \& Warner 1998), and V1974 Cyg ($P_{orb}$ = 1.950 h, $P_{sh}$ = 2.039 h: Retter, Leibowitz \& 
Ofek 1997).

For a disc inclination of 65$^{\circ}$, the correction in magnitude to the standard inclination of 57$^{\circ}$
is $-0.4$ (Warner 1995). At minimum light RS Car would therefore appear at $V$(min) $\approx 18.1$ at the standard
inclination. For a nova with $t_2 = 32$ d and standard inclination, the range of eruption should be $\sim$13.1 (Figure
5.4 of Warner 1995) which would make RS Car $V \sim 5.0$ at maximum, which adds support to the 
suspicion that maximum light was missed (Duerbeck 1987).

\subsection{V365 Carinae}

V365 Car was Nova Carinae 1948, a very slow nova reaching $m_{pg} = 10.1$ at maximum. It is listed in the Downes, Webbing \& Shara
(1997) catalogue as $m_j = 21.6$ at minimum, but photometry by Zwitter \& Munari (1996) in 1995 gave $V = 18.31$, $U-B = -0.51$
and $B-V = 0.63$, and Downes \& Duerbeck (2000) give $V = 18.47$ in 1998. The eruption range of $\sim 9$ mag is normal for the speed 
class. The spectrum obtained by Zwitter \& Munari shows strong HeII and moderate Balmer emission; the slope of the continuum is
shallow because of reddening of this distant low Galactic latitude object. There is no X-ray detection.

\begin{figure}
\centerline{\hbox{\psfig{figure=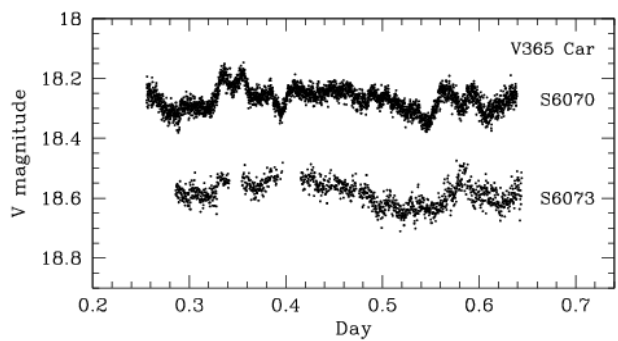,width=8.2cm}}}
  \caption{The light curves of V365 Car obtained in 2000 March. The light curve of run S6073 is displaced 
vertically by 0.4 mag for display purposes.}
 \label{lcv365car}
\end{figure}

Our high speed photometric runs on V365 Car are listed in Table 1 and the two longest light curves, spaced two days apart, 
are presented in Fig.~\ref{lcv365car}. The
Fourier transform of these light curves shows no features other than the fundamental and harmonics associated with the
slight upward convexity in the two long runs and the slow flickering. However, the two light curves seen in Fig.~\ref{lcv365car}
show a strong resemblence to each other, which could be the result of a modulation period that is near to a submultiple
of the two day spacing, e.g., 6.86 h or 8.0 h. These light curves illustrate a limitation of our
photometric survey: despite two long runs of good quality of $\sim$9 h each, no definitive modulation was
detected, and it was deemed not good use of available time to continue on this star. In effect, we are insensitive to low
amplitude modulations with periods $\ga$ 8 h.

\subsection{V436 Carinae}

V436 Car was identified as the optical counterpart ($V \sim 15.3$) of the ROSAT source RX\,J0744.9-5257, having strong
Balmer, moderate HeI and weak HeII emission lines, together with evidence for long term variations in brightness by $\sim 2$ mag
(Motch et al.~1996). Ramsay, Buckley \& Cropper (1998) obtained time-resolved spectroscopy and photometry of V436 Car and deduced
a probable orbital period of 3.60 h from the former and a possible 2.67 h modulation in the latter. No X-ray periodic modulation
was detected, nor any optical polarization. From its X-ray luminosity and the X-ray/optical ratio, Ramsay et al.~(1998) suggested
that V436 Car may be an intermediate polar, with rotational and magnetic axes nearly aligned.

V436 Car currently has no clear interpretation: its spectrum resembles that of a dwarf nova in quiescence, but 
its X-ray properties are unlike a dwarf nova and are more nearly those of an intermediate polar.

\begin{figure}
\centerline{\hbox{\psfig{figure=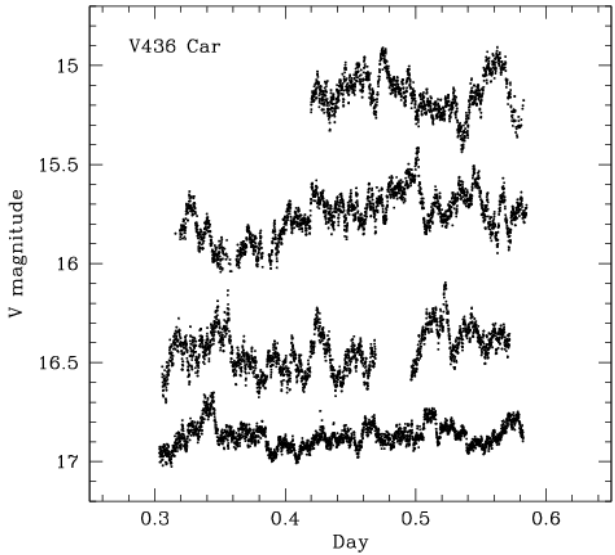,width=8.2cm}}}
  \caption{The light curves of V436 Car. The upper light curve reflects the correct brightness of V436 Car. The remaining three light curves
have been displaced vertically for display purposes by 0.55 mag, 1.2 mag 
and 2.35 mag, respectively, from top to bottom.}
 \label{lcv436car}
\end{figure}

\begin{figure}
\centerline{\hbox{\psfig{figure=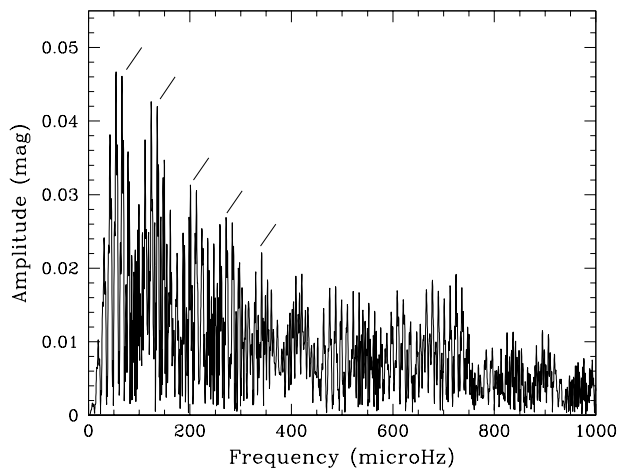,width=8.2cm}}}
  \caption{The Fourier transform of the combined runs S6039, S6040, S6041 and S6044 on V436 Car in 2000 February.
The slanted markers show the fundamental, and first four harmonics, of the 4.207 h periodic signal in V436 Car.}
 \label{ftv436car}
\end{figure}

Our photometric observations on V436 Car are listed in Table 1 and the light curves are illustrated in Fig.~\ref{lcv436car}. 
The Fourier transform of the combined runs S6039, S6040, S6041, and S6044, is shown in Fig.~\ref{ftv436car} and is clearly
dominated by a periodic signal with an extensive set of harmonics. The period that we find to be most compatible with 
the aliases of the fundamental and harmonics is $P_{orb}$ = 4.207 h. This corresponds closely with the 0.1761 d (= 4.226 h)
period reported by Ramsay et al.~(1998) to be a strong alias of their spectroscopic period. There is no other convincing
coincidence  of aliases in the spectroscopic and photometric Fourier transforms. We conclude that our photometric
observations have confirmed the results obtained from the spectroscopic study.

\begin{figure}
\centerline{\hbox{\psfig{figure=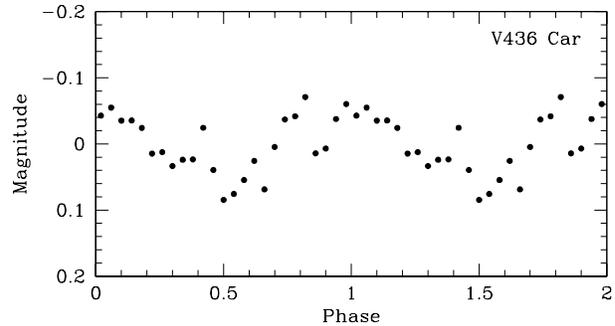,width=8.2cm}}}
  \caption{The average light curve of V436 Car, folded on the 4.207 h period.}
 \label{meanv436car}
\end{figure}

\begin{figure}
\centerline{\hbox{\psfig{figure=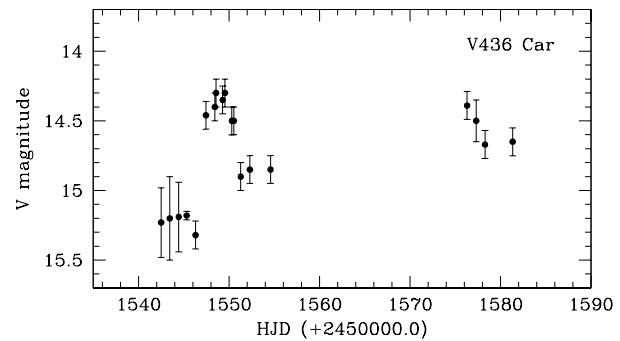,width=8.2cm}}}
  \caption{The long term light curve of V436 Car in 2000 January and February. The errorbars indicate
the observed range of variability per night.}
 \label{ltv436car}
\end{figure}

The mean light curve at the 4.207 h period is shown in Fig.~\ref{meanv436car} and has a range of $\pm$0.08 mag.
Our observations in 2000 January showed that V436 Car had brightened by $\sim$0.75 mag on the final night
of our run (S6044: the amplitude of the fast variations is noticeably reduced by the addition of the constant
background, see Fig.~\ref{lcv436car}). In order to investigate the nature of this brightening, the observer on the telescope
during the following week -- Dr.~E.~Romero Colmenero -- kindly obtained some short runs. The long term light
curve is shown in Fig.~\ref{ltv436car} and shows a rise of about 0.8 mag followed by a decay two days later. The speed of
the initial rise (0.75 mag in less than a day) and the time scale of the decay are compatible with a dwarf nova
in outburst, but the pause during decline is not, so more frequent monitoring of V436 Car is 
required in order to settle its CV subtype.

The harmonics of $P_{orb}$ seen in Fig.~\ref{ftv436car} coincide in some cases with the structures in the Fourier
transform of the photometric data obtained by Ramsay et al.~(1998). Their marked peaks at 62 min and 52 min are
the third and fourth harmonics, respectively, and the first harmonic is probably present. The fundamental is not 
clearly visible in their work because of the dominance of runs with length less than, or about equal to, $P_{orb}$.

In addition, Ramsay et al.~find considerable power near 24 min and 160 min. The latter they interpret as possibly
the signature of an intermediate polar with spin or reprocessing period of 2.67 h. 

As can be seen from the Fourier transform of the total data set (Fig.~\ref{ftv436car}), we do not see any persistent
modulations at these or other periods (though there is a broad feature centred on $\sim$400 $\mu$Hz $\approx$ 42 min).
On individual nights, however, there are significant peaks in the Fourier transforms (e.g., $\sim$25 min in S6039, 
$\sim$14.8 min and $\sim$23 min in S6040, $\sim$27.3 min in S6041 and $\sim$30.5 min in S6044). These are
characteristic of the quasi-periodic oscillations (QPOs) seen in CVs (e.g., Woudt \& Warner 2002). There is also
a significant band of power at $\sim$123 s in run S6040 and dwarf nova oscillations (DNOs) at $\sim$40 s in the second
half of run S6039. These QPOs and DNOs will be analysed in detail elsewhere.

It should be noted that no intermediate polar has been seen to possess DNOs, and that this is a natural
consequence of the magnetic accretion model for DNOs (Warner \& Woudt 2002). 

We conclude that there is no evidence that V436 Car is an intermediate polar.

\subsection{AP Crucis}

AP Cru was Nova Crucis 1936, reaching $m_{pg} = 10.7$, and is listed in the Downes et al.~(1997) catalogue
at $m_j = 21.7$. There were no spectra obtained during eruption and no detailed light curve was obtained, which
leaves the classification of AP Cru uncertain -- even its status as a nova is unclear, though the eruption 
range of 11 mag is strongly suggestive. Munari \& Zwitter (1998) obtained a spectrum that shows AP Cru to be
uncharacteristic of CVs: it has an absorption spectrum like a K7 star with quite strong and very narrow
H$\alpha$ emission superimposed. They also obtained $V = 18.68$, $B-V = 1.37$ and $U-B = 0.60$. There is no
X-ray detection of AP Cru.

\begin{figure}
\centerline{\hbox{\psfig{figure=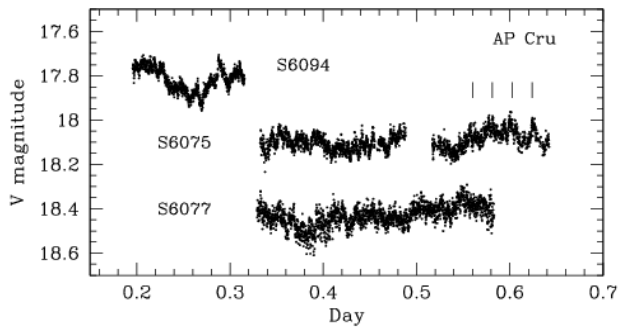,width=8.2cm}}}
  \caption{The light curves of AP Cru in 2000 March and June. The light curve of run S6077 has been shifted vertically
by 0.4 mag for display purposes. The vertical bars indicate prominent maxima in run S6075 of the 1837 s modulation.}
 \label{lcapcru}
\end{figure}

Our four high speed photometric runs are listed in Table 1 and the light curves are shown in Fig.~\ref{lcapcru}. In
the three longer runs there is evidence for modulation on a time scale of $\sim$ 2.5 hours. Combining the two long runs made
in 2000 March produces the Fourier transform shown in Fig.~\ref{ftapcru}, where the mean magnitude and trend has been
removed from the individual light curves. From this we obtain a period of 2.56 h.

\begin{figure}
\centerline{\hbox{\psfig{figure=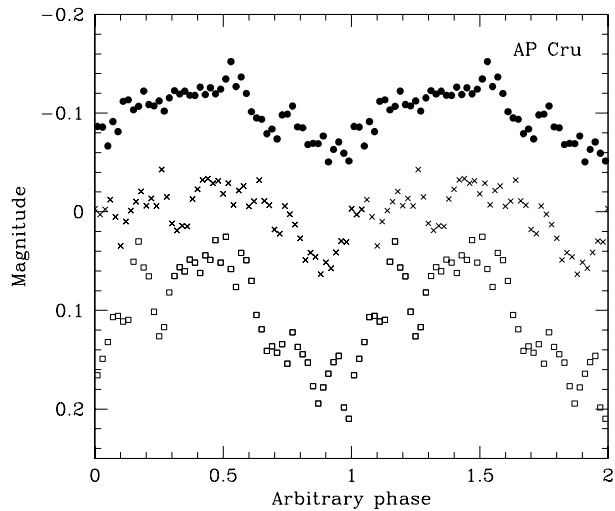,width=8.2cm}}}
  \caption{The mean light curves of AP Cru for runs S6075 (filled circles), S6077 (crosses), and S6094
(open squares), respectively. The mean light curves of runs S6075 and S6094 have been displaced by -0.1 and +0.1 mag, 
respectively, for display purposes. The individual light curves have been aligned in phase.}
 \label{meanorbapcru}
\end{figure}

\begin{figure}
\centerline{\hbox{\psfig{figure=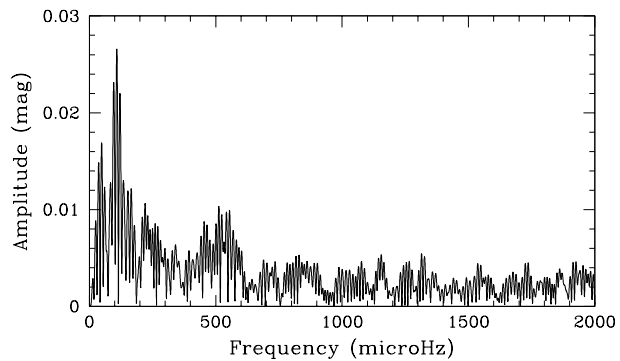,width=8.2cm}}}
  \caption{The Fourier transform of the two combined long runs on AP Crucis in 2000 March.}
 \label{ftapcru}
\end{figure}

AP Cru was 0.6 -- 0.9 mag brighter during our photometric runs than when Munari \& Zwitter (1998) obtained their spectrum. 
We suggest that the star was in a state of very low or zero rate of mass transfer then, revealing the K7 secondary with 
H$\alpha$ emission from its chromosphere (heated by radiation from the primary). The additional luminosity present during
our observations arises from an accretion disc produced by the resumption of mass transfer. However, the additional
luminosity is only about equal to that of the secondary, so the flux from AP Cru (especially in the red, where our CCD
is most sensitive) will continue to be dominated by that from the secondary. Consequently, we expect that our
observed 2.56 h modulation is caused by ellipsoidal variation of the secondary, and therefore the orbital period is
5.12 h. The latter period is more compatible with the Spectral Type -- $P_{orb}$ correlation for CVs (see Figure 1
of Smith \& Dhillon 1998) than is 2.5 h. The mean light curves for each long run, calculated from the 2.56 h periodicity,
are shown in Fig.~\ref{meanorbapcru}.

\begin{figure}
\centerline{\hbox{\psfig{figure=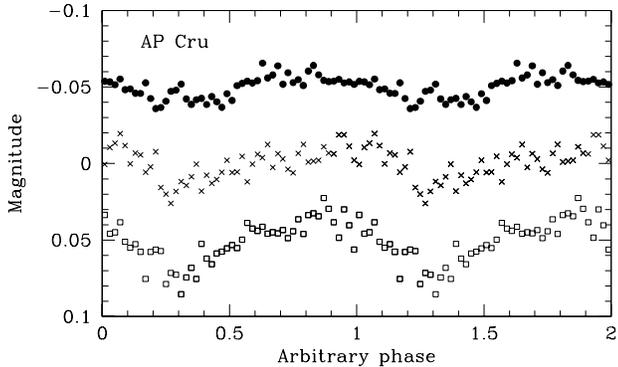,width=8.2cm}}}
  \caption{The mean light curves of AP Cru at the 1837 s modulation for runs S6075, S6077, and S6094, respectively. 
The symbols are as in Fig.~\ref{meanorbapcru}. The mean light curves of runs S6075 and S6094 have been displaced by -0.05 
and +0.05 mag, respectively, for display purposes. The individual light curves have been aligned in phase.}
 \label{meanspapcru}
\end{figure}

The Fourier transform (Fig.~\ref{ftapcru}) shows a significant band of power around 500 $\mu$Hz. On examining
this in detail, we discovered that it is due to a coherent modulation, broadened by amplitude modulation.
The signal is present in all four of our light curves and has a period of 1837 s (30.61 min). Some maxima of this modulation
in run S6075 are marked in Fig.~\ref{lcapcru}. The mean light curves at 1837 s for our three long runs are shown in 
Fig.~\ref{meanspapcru}; the amplitude is noticeably larger in S6094 when the system was brighter.

\begin{figure*}
\centerline{\hbox{\psfig{figure=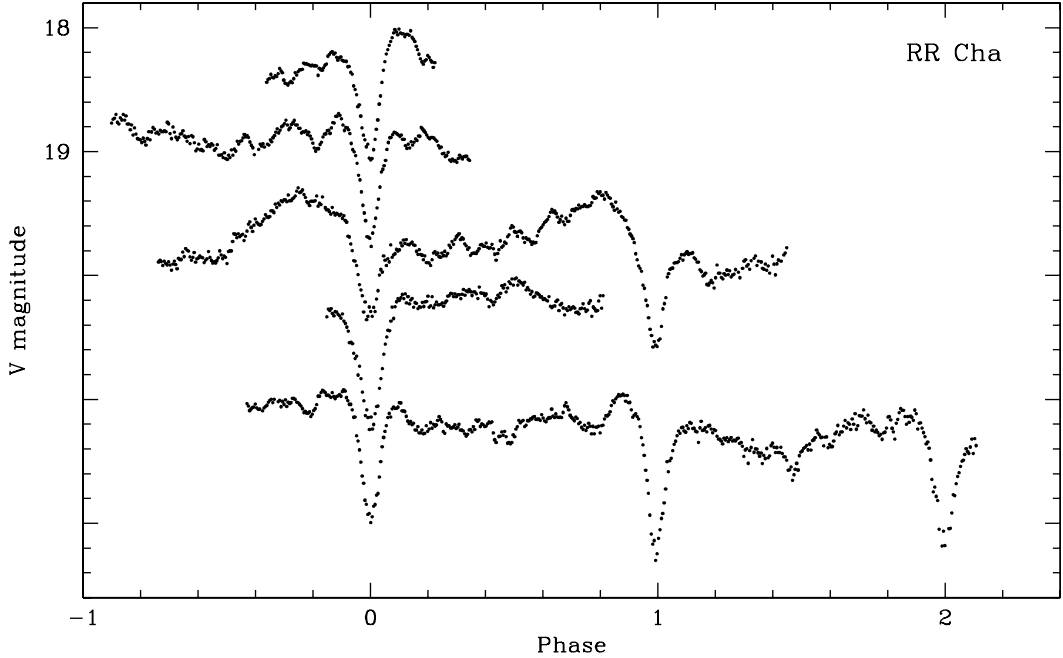,width=16cm}}}
  \caption{The light curves of RR Cha, phased according to the ephemeris in Eqn.~\ref{ephrrcha}.
The upper light curve reflects the correct brightness of RR Cha. Subsequent light curves are shifted 
vertically by 0.6 mag, 1.2 mag, 1.9 mag, and 2.7 mag, respectively, from top to bottom, for display purposes. }
 \label{lcrrcha}
\end{figure*}

A persistent signal of this kind is characteristic of an intermediate polar. Unfortunately, AP Cru is at low
Galactic latitude (2$^{\circ}$) and at a large distance, so it is unlikely to be detectable in the X-ray
region ($P_{orb} \approx 5$ h gives $M_V \sim 10.3$ for the secondary (Warner 1995); combined with the apparent
magnitude at the time of Munari \& Zwitter's observation this gives a distance of $>$ 500 pc even without
allowance for reddening).

\subsection{RR Chamaeleontis}

RR Cha reached at least $m_{pg} = 7.1$ at maximum as Nova Chamaeleontis 1953. It was a moderately fast 
nova and is listed in the Downes et al.~(1997) catalogue at $m_j = 19.3$. Gill \& O'Brien (1998) have
detected a nova shell around RR Cha. Zwitter \& Munari (1996) found $V = 18.93$ and obtained a
spectrum showing strong HeII and Balmer emission characteristic of a hot post-nova.

Our high speed photometric observations of RR Cha are listed in Table 1 and the light curves appear in
Fig.~\ref{lcrrcha}. It can be seen that RR Cha is an eclipsing system with large humps that do not
remain locked in orbital phase -- they appear to be superhumps. The varying depth of the eclipses 
seen in Fig.~\ref{lcrrcha} is another characteristic of a superhumping system.

The orbital period determined from the eclipses in the three nights of 2001 February, is
3.362 h. The gap between the 2001 February (group 1) and 2001 May (group 2) observations is too
large to bridge in order to get a more precise period. The ephemeris derived from group 1 is

\begin{equation}
  {\rm HJD_{min}} = 2451965.5964 + 0.1401 {\rm E}
 \label{ephrrcha}
\end{equation}

\begin{figure}
\centerline{\hbox{\psfig{figure=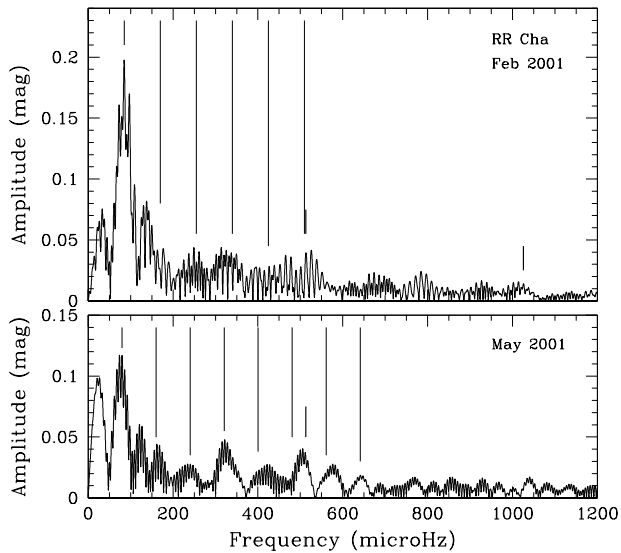,width=8.2cm}}}
  \caption{The Fourier transforms of the 2001 February (upper panel) and the 2001 May (lower panel)
observations of RR Cha. The eclipses have been removed from the light curves. The thin vertical
bars point to the fundamental and harmonics described in the text. The thick bars indicate the 1950 s
modulation and its first harmonic.}
 \label{ftrrcha}
\end{figure}

Because of the presence of narrow eclipses, Fourier transforms of the light curves of RR Cha are
dominated by the fundamental and harmonics of the orbital period. We therefore removed the eclipses from
the light curves and the resulting Fourier transforms are shown in Fig.~\ref{ftrrcha}.

We will start by describing the Fourier transform of the combined two nights of group 2 (lower panel
of Fig.~\ref{ftrrcha}). Removal of the orbital harmonics (as a result of excising the eclipses) reveals
a harmonic structure on the lower frequency wings of the orbital harmonics. This is due to a non-sinusoidal
modulation with a period of 3.466 h (80.1 $\mu$Hz). 
The peak at 120.9 $\mu$Hz  we suspect
is an artefact due to a short data length in one of the observing runs. The 3.466 h modulation is probably a
positive superhump, with a period 3.1\% longer than the orbital period.

A striking feature of the Fourier transform is the presence of a strong modulation at a period near 1970 s.
This is far removed from a harmonic of either the orbital or the superhump periods. There are three aliases
in this peak: 1954 s, 1976 s and 2000 s. The first harmonic of this modulation is only just visible above the noise
in the Fourier transform.

In the Fourier transform of the three nights of group 1 (upper panel of Fig.~\ref{ftrrcha}), the dominant feature
is the window pattern centred on 3.271 h. This is almost purely sinusoidal, the amplitude of the harmonics is very
low. This is evidently a negative superhump, with a period 2.7\% shorter than the orbital period. Because the superhump
amplitude varies, the harmonics have additional components caused by amplitude modulation.

In group 1, the $\sim$1970 s modulation again appears strongly, with three aliases: 1859 s, 1903 s and 1948 s. 
We are able to select among these aliases by appeal to the first
harmonic of the $\sim$1970 s modulation, where we find 964 s, 976 s and 988 s, respectively. This determines 
a unique solution, i.e.~1950 s = 32.5 min. This peak coincides with the 1954 s alias in the 2001 May observations.
In the upper panel of Fig.~\ref{ftrrcha} it can be seen that the 5th harmonic of the negative superhump, if it were present,
would be close to the 1950 s 
peak. They can, however, be clearly distinguished from each other: the window pattern is of the 1950 s modulation, the
5th harmonic would lie at a minimum of this pattern. The higher amplitude of the 1950 s modulation is at times 
very obvious in the light curve, see Fig.~\ref{lcrrcha}. 

The high stability of the 1950 s modulation excludes the possibility of quasi-periodic oscillations of the
kind often seen in CVs (e.g.~VW Hyi: Woudt \& Warner 2002; Warner \& Woudt 2002). This modulation and the high
ionization emission lines seen in the spectrum, mentioned above, point to RR Cha being an intermediate polar (e.g.~Warner 1995).
RR Cha is not detected in the ROSAT survey (G. Israel, private communication), which is not unexpected given the
high inclination with resulting obscuration of X-rays by the disc.
We conclude that RR Cha is an eclipsing old nova in which both positive and negative superhumps occur, and furthermore
it is an intermediate polar. We cannot tell from the Fourier transforms whether the 1950 s modulation is the spin
period of the white dwarf or a reprocessed period. RR Cha offers great potential for examining the interaction
between the rotating beam of radiation from the white dwarf and the structures (e.g.~elliptical prograde precessing disc, 
and/or retrogradely precessing tilted disc) in the system.

\subsection{BI Orionis}

BI Ori is classified as a dwarf nova, perhaps of Z Cam type, with a listed range $V$ = 13.2 -- 16.7. Bateson, McIntosh \& Stubbings
(1997) find a mean outburst interval of 22 d and two types of outburst. No orbital
period is known and only $\sim$ 2 h of high speed photometry has been published (Szkody 1987), which shows flickering
at minimum light. The spectrum at quiescence shows the strong Balmer emission characteristic of a dwarf nova;
the lines are broad and possibly double, indicative of a system of moderate inclination (Szkody 1987). BI Ori is
a weak ROSAT source (Richman 1996).

Our observations of BI Ori are listed in Table 1. The brightening on 2001 October 11 was also recorded by the AAVSO, as an 
isolated observation. Our observations show that it was still bright on October 15, which puts it into the long outburst
duration category of Bateson et al.~(1997).

\begin{figure}
\centerline{\hbox{\psfig{figure=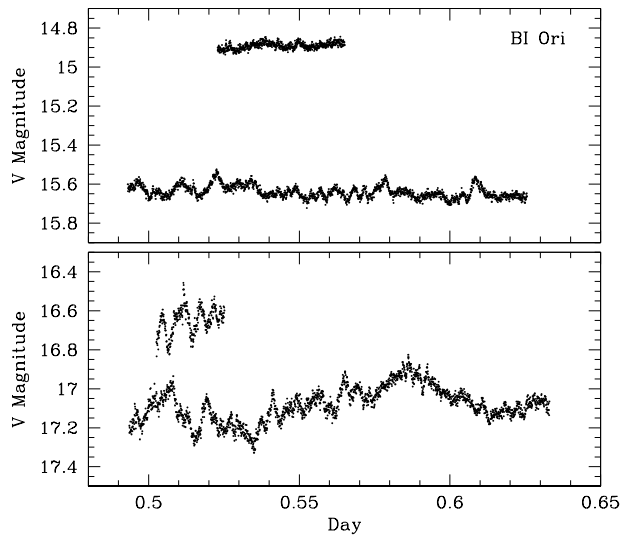,width=8.2cm}}}
  \caption{The light curves of BI Ori in 2001 September and October. The light curve of run S6261 has 
been displaced vertically downwards by 0.7 mag for display purposes.}
 \label{lcbiori}
\end{figure}

The high speed photometry shown in Fig.~\ref{lcbiori} shows the flickering typical of a dwarf nova in quiescence and 
its reduction in amplitude during outburst. In none of the light curves is there a hint of an orbital modulation.
Fourier transforms of the light curves reveal no short time scale periodicities.

\subsection{CM Pheonix}

CM Phe, designated Phe1 in the Downes et al.~(1997) catalogue, is a nova-like variable which was for many
years misidentified until recovered by Jaidee \& Lyng\^a (1969) and confirmed to be a rapidly flickering
object by Koen \& O'Donoghue (1995). Hoard \& Wachter (1998) found $V$ = 15.3 and confirmed rapid photometric
variability. In a medium-resolution spectrum they found strong H$\alpha$ and moderate strength He\,I emission
on a blue continuum, but also detected TiO bands from a secondary component estimated to have spectral
type M2--5.

The existing high speed photometry being of short duration and limited interpretational value, we added
CM Phe to our candidate list. The observations we made in December 2000 were barely sufficient to give an unambiguous
determination of orbital modulation, but in the meantime a more extensive study has been published by
Hoard, Wachter \& Kim-Quijano (2001; hereafter HWK), so we terminated our own work and use it here only
to supplement that of HWK.

\begin{figure}
\centerline{\hbox{\psfig{figure=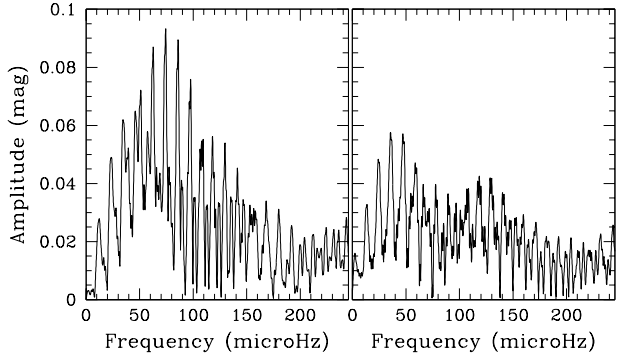,width=8.2cm}}}
  \caption{The Fourier transform of all our observations of CM Phe (left panel). The right panel shows the Fourier
transform prewhitened at the highest peak of the left panel, i.e.~at 74.4 $\mu$Hz.}
 \label{ftcmphe}
\end{figure}

Our observing runs are listed in Table 1. For the last two of our runs we used the SAAO CCD photometer
instead of the UCT CCD photometer (which we used simultaneously on the 1.9-m telescope). The data taken with
the SAAO CCD were uncalibrated. The Fourier
amplitude spectrum of the whole data set is shown in Fig.~\ref{ftcmphe}. Guided by the conclusion of HWK, namely
that the light curve has two unequal maxima and minima in each cycle, we list candidates for the fundamental, first
and second harmonics (ordered by amplitude at each harmonic) in Table~\ref{tab2}.

\begin{table}
\caption{Candidates for the fundamental, first and second harmonics in CM Phe.}
\begin{tabular}{lll}
                & P (days) & P (days) \\[10pt]
Fundamental     & 0.325  & 0.245  \\
First harmonic  & 0.1556 & 0.1345 \\
Second harmonic & 0.0976 & 0.0900  \\
\end{tabular}
\label{tab2}
\end{table}

We have quoted the periods at the maxima of the peaks (in the Fourier transform) and their one-day aliases,
without formal error estimates, which are difficult to estimate realistically. Looking first at the first 
harmonic, our values are identical within errors to the values 0$\fd$1552 (21) and 0$\fd$1348 (17) found by
HWK. For the fundamental HWK deduced 0$\fd$2667 (62) as the most probable, on the grounds that it is close 
to twice the period of the 0$\fd$1348 1st harmonic, and that they found no significant amplitude at 2 $\times$
0$\fd$1552. In our case, our 0$\fd$325 candidate for the fundamental is closer to 2 $\times$ 0$\fd$1556 than
0$\fd$245 is to 2 $\times$ 0$\fd$1345. However, appeal to the second harmonic produces an unambiguous 
choice.

We note first that the cleaned power spectrum produced by HWK has a peak at 11.2 d$^{-1} \equiv 0\fd089$, i.e.
${1\over{3}} \times 0\fd2667$, but not at the 2nd harmonic of any 1 day$^{-1}$ alias of 0$\fd$2667. Their
2nd harmonic matches ours of 0$\fd$0900; both are close to ${2\over{3}}$ of our well determined first
harmonic at 0$\fd$1556. On the other hand, our alias at 0$\fd$0976 is not close to ${2\over{3}}$ of either
of our candidate values of the first harmonic.

In conclusion, we confirm HWK's choice of aliases: the fundamental period is 2 $\times$ 0$\fd$1345 = 0$\fd$2690.
With our baseline of 9 d, compared to HWK's 5d, we expect our determination of the period to be of slightly 
higher precision. A non-linear least squares fit of a sinusoid to the first harmonic in our data provides
the following ephemeris for the times of the deepest of the minima:

\begin{equation}
  {\rm HJD_{min}} = 2451900.4561 +  0.2689 (\pm 7) {\rm E}
 \label{ephcmphe}
\end{equation}

\begin{figure}
\centerline{\hbox{\psfig{figure=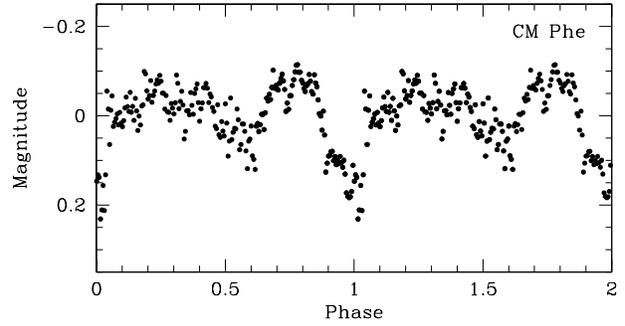,width=8.2cm}}}
  \caption{The average light curve of CM Phe, folded on the 0.2689 day period. Each bin contains the average 
of about 40 measurements.}
 \label{cmphe_av}
\end{figure}

The mean light curve folded on this period is shown in Fig.~\ref{cmphe_av}.

\begin{figure}
\centerline{\hbox{\psfig{figure=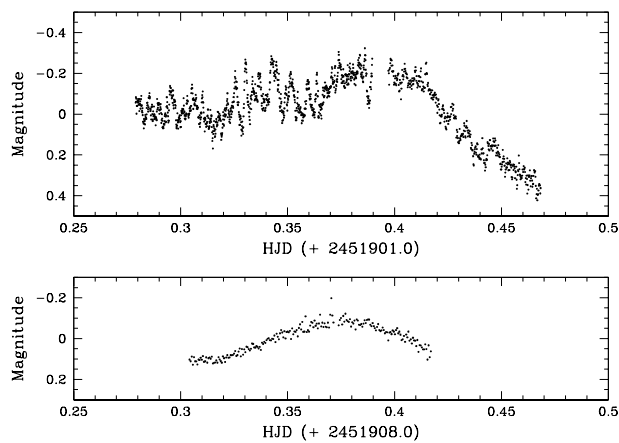,width=8.2cm}}}
  \caption{The upper panel shows the light curve of CM Phe taken on 2000 Dec 22 with the UCT CCD. The lower panel
displays the light curve of CM Phe on 2000 Dec 29, obtained with the SAAO CCD. Both observations were made in white light.}
 \label{cmphe_ex}
\end{figure}

In the shorter period domain we have found no significant modulations below periods of $\sim$500 s. There is, however,
an interesting difference in the flickering amplitude as registered by the UCT CCD photometer and the SAAO
CCD photometer. The former has good sensitivity in the blue and near UV, whereas the latter has high red and near-infrared
sensitivity but very low short wavelength efficiency. All of our light curves obtained with the UCT instrument
shows high speed activity -- an example is given in the upper panel of Fig.~\ref{cmphe_ex}. In contrast,
all the light curves obtained with the SAAO instrument are smooth and apparently free of rapid activity -- for an example see lower
panel of Fig.~\ref{cmphe_ex}. We consider this to be in accord with the conclusion of HWK that the double wave
modulation is due to ellipsoidal variation of the red secondary, to which is added accretion-related flickering
that has a higher energy flux distribution.

\subsection{V522 Saggitarii}

V522 Sgr was discovered at $m_{pg} \sim$ 12.9 in August 1931 and faded to $V \sim 17.5$ (Duerbeck 1987;
Downes et al.~1997). The small amplitude is uncharacteristic of a nova, leaving the possibility that V522 Sgr
is a dwarf nova (Duerbeck 1987). However, the spectrum of the candidate nova remnant is that of a G2 -- K5 star
(Ringwald, Naylor \& Mukai 1996), throwing doubt on the correctness of its identification.

\begin{figure}
\centerline{\hbox{\psfig{figure=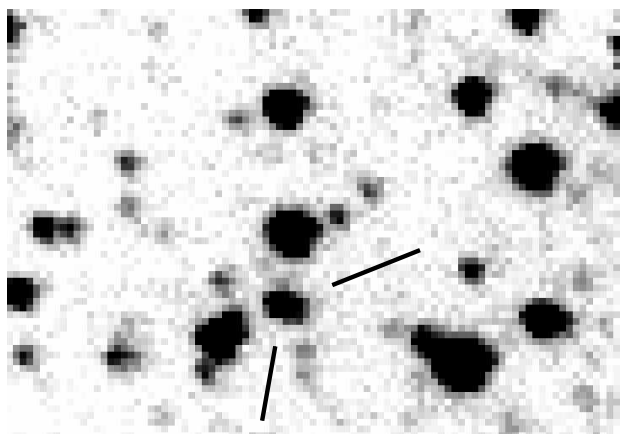,width=8.2cm}}}
  \caption{Finding chart for V522 Sgr. The variable is marked with bars.
The field of view is 50 by 34 arcsec, north is up and east is to the left.}
 \label{fcv522sgr}
\end{figure}

We obtained one high speed photometric run centred on the nova candidate (Table 1). The candidate and all but one
of the other stars in the vicinity showed no flickering activity. The only non-constant star is 6 seconds of arc 
south of the candidate (a finding chart is given in Fig.~\ref{fcv522sgr}) at $V \sim$ 19.0, and showed some slow
low amplitude flickering in the last half of our run. We suggest this to be the true remnant of Nova
Saggitarii 1931. However, the image is elongated which suggests it is a blend of two stars each of magnitude $\sim$ 19.7.
This would give a revised eruption amplitude of $\sim$ 6.8 mag which is still compatible with a long interval outbursting
dwarf nova of the WZ Sge type.

\section{Conclusions}

We have presented high speed photometry for eight faint CVs, of which five are old novae. Periodic modulations are definitely
detected in three of the five novae, giving orbital periods for two (AP Cru and RR Cha) and a probable superhump
period for RS Car. In addition, a possible long period modulation of undetermined period is found in V365 Car. An alternative
candidate is proposed from the remnant of Nova Saggitarii 1931.

RR Cha is found also to be an intermediate polar candidate, with a period of 32.5 min (which could be either the spin
period of the white dwarf or its orbital sideband). RR Cha is therefore a deeply eclipsing intermediate polar -- and in 
addition it shows positive superhumps and negative superhumps.

AP Cru is also an intermediate polar candidate, with an observed period of 30.61 min

The nova-like variable V436 Car has a photometric period of 4.207 h, which is an alias of the spectroscopic 
period previously found by Ramsay et al.~(1998) but does not show the 2.67 h modulation which led those authors
to propose its intermediate polar candidature. Instead, we find it to have DNOs near 40 s, which is compatible with
previous findings that intermediate polars and DNOs do not co-exist.

For BI Ori we find no orbital modulation in the range of periods accessible to us.

From our photometry of CM Phe, made without knowledge of that being carried out by HWK, we concur with their choice 
of $P_{orb}$ from among the various harmonics and aliases.

\section*{Acknowledgments}
We thank Dr. D.~O'Donoghue for the use of his EAGLE program for Fourier analysis of the light curves. 
Dr. G. Israel kindly searched in the ROSAT database at our request. We kindly acknowledge Dominic de Guzman.
PAW is funded partly through strategic funds made available to BW by the University of Cape Town. BW's research
is funded entirely by that university.

\end{document}